# A new model of dark matter distribution in galaxies

Dragan Slavkov Hajdukovic[1,2]
[1]Institute of Physics, Astrophysics and Cosmology; Cetinje, Montenegro
[2]Physics Department, CERN; CH-1211 Geneva 23
dragan.hajdukovic@cern.ch

Abstract
In the absence of the physical understanding of the phenomenon, different *empirical* laws have been used as approximation for distribution of dark matter in galaxies and clusters of galaxies. We suggest a new profile which is not empirical in nature, but motivated with the physical idea that what we call dark matter is essentially the gravitational polarization of the quantum vacuum (containing virtual gravitational dipoles) by the immersed baryonic matter. It is very important to include this new profile in forthcoming studies of dark matter halos and to reveal how well it performs in comparison with empirical profiles. A good agreement of the profile with observational findings would be the first sign of unexpected gravitational properties of the quantum vacuum.

## 1. Introduction

In galaxies and clusters of galaxies, *the gravitational field is much stronger* than it should be according to our theory of gravitation and the existing amount of baryonic matter. No one knows how to explain this phenomenon. The prevailing *hypothesis* is the existence of a huge quantity of dark matter in the Universe. Of course it is possible that dark matter of unknown nature exists, but it is also possible that it is only a theoretical construct that *mimics well, something quite different*. A recent suggestion (Hajdukovic 2011, 2012) is that with dark matter, we *mimic* phenomena caused by *baryonic matter immersed in the quantum vacuum*. The new theory uses the quantum vacuum well established (Aitchison, 2009) in quantum field theory, enriched with *only one additional feature*: it contains *virtual gravitational dipoles*. Within this framework, the gravitational polarization of the quantum vacuum (analogous to polarization of a dielectric in external electric field, or a paramagnetic in an external magnetic field) produces effects as if there were an *effective* quantity of dark matter.

In the absence of physical understanding of the phenomenon, the distribution of (real or *effective*?) dark matter in a galaxy is usually described by empirical laws (NFW profile, Einasto profile, Burkert profile...). For instance the Burkert profile is:

$$\rho_B(r) = \frac{\rho_0}{\left(1+\frac{r}{r_0}\right)\left[1+\left(\frac{r}{r_0}\right)^2\right]} \qquad (1)$$

where $\rho_0$ (central density) and $r_0$ (characteristic radius) are free parameters determined to give the best possible fit of observational findings for considered galaxies.

In the present paper we suggest a new distribution giving directly the quantity of *effective* dark matter within a sphere of radius $r$:

$$M_{dm}(r) = 4\pi r^2 P_{g\,max} \tanh\left(\frac{R_c}{r}\right) \qquad (2)$$

with corresponding density distribution

$$\rho_{dm}(r) = \frac{dM_{dm}(r)}{dV} = \frac{P_{g\max}}{r}\left[2\tanh\left(\frac{R_c}{r}\right) - \frac{R_c}{r}\frac{1}{\cosh^2\left(\frac{R_c}{r}\right)}\right] \quad (3)$$

While constants $P_{g\max}$ and $R_c$ in equations (2) and (3) can be roughly estimated (in the framework of the theoretical model of the quantum vacuum enriched with virtual gravitational dipoles) it would be better to consider them as free parameters in the same way as $\rho_0$ and $r_0$ in empirical law (1).

The key purpose of this short article is to bring the relation (2) to the necessary attention of astronomers specialized in the study of dark matter distribution, who can confront (2) with observations. Relation (2) having physical roots is potentially much more significant than the proposal of any additional empirical law would be. In our opinion it is both urgent and important to include relation (2) in forthcoming studies of dark matter distribution in different halos and to reveal how good its performance is in comparison with empirical profiles. For instance, in very recent publications, different empirical profiles have been compared with observational findings for the Milky Way (Boylan-Kolchin et al. 2013; Nesti, 2013) and Andromeda galaxy (Tamm et al. 2012); in future studies of such a kind, distribution (2) must be considered. A good agreement of profile (2) with observational findings would be the first sign of unexpected gravitational properties of the quantum vacuum and a challenge to the existence of dark matter.

In Section 2 we present theoretical arguments for distribution (2), while Section 3 is devoted to discussions. In Appendix we argue that the observed dark matter halo surface density in galaxies (Kormendy and Freeman 2004, Spano et al. 2008, Donato et al. 2009) might be interpreted as a signature of quantum vacuum (i.e. distribution (2))

## 2. Theoretical background of the new distribution

According to previous papers (Hajdukovic 2011, Hajdukovic 2012), if antiparticles have negative gravitational charge, a virtual particle-antiparticle pair in the quantum vacuum can be considered as a gravitational dipole, with the gravitational dipole moment

$$\vec{p}_g = m_g \vec{d}, \quad |\vec{p}_g| < \frac{\hbar}{c} \quad (4)$$

Here, by definition, the vector $\vec{d}$ is directed from the antiparticle to the particle, and has a magnitude $d$ equal to the distance between them. The inequality in (4) follows from the fact that the size $d$ of the virtual pair must be smaller than the reduced Compton wavelength $\lambdabar_m = \hbar/mc$ (for a larger separation a virtual pair becomes real). Hence, $|\vec{p}_g|$ must be a fraction of $\hbar/c$.

Consequently, a gravitational polarization density $\vec{P}_g$ (i.e. the gravitational dipole moment per unit volume) may be attributed to the quantum vacuum. The spatial variation of the gravitational polarization density $\vec{P}_g$ generates an effective gravitational charge density, $\rho_{dm} = -\nabla \cdot \vec{P}_g$ which apparently has the potential to explain the observed phenomena without invoking dark matter. In the case of spherical symmetry (Hajdukovic 2011, Hajdukovic 2012)

$$\rho_{dm}(r) = \frac{dM_{dm}}{dV} = \frac{1}{r^2}\frac{d}{dr}\left(r^2 P_g(r)\right) \quad (5)$$

and consequently

$$M_{dm}(r) = 4\pi r^2 P_g(r) \tag{6}$$

The behavior of function $P_g(r)$ is known (Hajdukovic 2011) in two limits

$$P_g(r \ll R_c) = P_{g\,max}, \quad P_g(r \gg R_c) = \frac{P_{g\,max} R_c}{r} \tag{7}$$

Our understanding of the quantum vacuum is not sufficient to find function $P_g(r)$ within the rigorous approach of quantum field theory. Fortunately, in spite of absence of detailed knowledge, we can get a crude approximation from consideration of an ideal system of non-interacting dipoles in an external gravitational field. Hence, we consider the gravitational polarization of the quantum vacuum as analogous to polarization of a dielectric in external electric field, or a paramagnetic in an external magnetic field. If so, paramagnetic ideal gas, ideal gas of electric dipoles and ideal gas of gravitational dipoles are three mathematically equivalent models.

It is well established in quantum mechanics that an electric dipole in an external electric field (or a magnetic dipole in an external magnetic field) can have only a finite number $n_e$ of different energies. In other words, the angle between the direction of the field and the direction of the dipole can have only $n_e$ different values. The same should be true for an ideal system of non-interacting gravitational dipoles.

The simplest physical case is $n_e = 2$ when dipoles can point only in two directions, in the direction of the field and opposite to the field; with the corresponding energy levels $-\varepsilon$ and $+\varepsilon$ which depend on $r$. In this case, the partition function for a system of $N$ non-interacting dipoles is

$$Z = \left(e^{\varepsilon/k_B T} + e^{-\varepsilon/k_B T}\right)^N = 2^N \cosh\left(\frac{\varepsilon(r)}{k_B T}\right) \tag{8}$$

where $k_B$ is the Boltzmann constant and $T$ the absolute temperature of the quantum vacuum.

From the partition function $Z$, it is easy to calculate the thermal average of the dipole moment and the corresponding polarization density

$$P_g(r) = P_{g\,max} \tanh\left(\frac{\varepsilon(r)}{k_B T}\right) \tag{9}$$

It remains to show that

$$\frac{\varepsilon(r)}{k_B T} = \frac{R_c}{r} \tag{10}$$

and then distribution (2) trivially follows from equations (6), (9) and (10).

From purely mathematical point of view, relation (10) follows from the demand that $P_g(r)$ determined by equation (9) has limits given by (7).

Of course physical reasons are always preferable to mathematical ones. While it will be subject of a forthcoming publication, let us give an initial physical argument that $k_B T$ and $R_c$ are inversely proportional as in equation (10). Physically there is a competition between energy $\varepsilon$ (favoring alignment of dipoles) and energy $k_B T$ (favoring randomness in orientation). The mean distance between two dipoles which are the first neighbors is $\lambda_m$. The gravitational acceleration produced by a particle of mass $m$ at the distance of its own Compton wavelength $\lambda_m$ is $Gm/\lambda_m^2$; in the absence of more accurate estimates, this acceleration can be used as a rough approximation (Hajdukovic 2011, Hajdukovic 2012) of the external gravitational field which is needed to produce the effect of saturation for the dipoles of mass $m$. In a similar way, the energy of gravitational interaction between two particles of mass $m$ at the distance of the corresponding Compton wavelength may be used as approximation for the energy $k_B T$, i.e. $k_B T \propto Gm^2/\lambda_m$. Hence $k_B T$ is inversely proportional to $\lambda_m$ and because of the direct proportionality (7) it is also inversely proportional to $R_c$.

For completeness, let us underline that equation (2) is a special case ($J = 1/2$) of

$$M_{dm}(r) = 4\pi r^2 P_{g\max} B_J\left(\frac{R_c}{r}\right) \tag{11}$$

where $B_J(x)$, with $J$ being a positive integer or half-integer, denotes Brillouin function

$$B_J(x) = \frac{2J+1}{2J}\coth\left(\frac{2J+1}{2J}x\right) - \frac{1}{2J}\coth\left(\frac{1}{2J}x\right) \tag{12}$$

Of notable interest, the Brillouin function is best known for arising in the calculation of the magnetization of an ideal paramagnet. In particular, it describes the dependency of the magnetization $M$ on the applied magnetic field $B$ and the total angular momentum quantum number $J$ of the microscopic magnetic moments of the material. In the analogous way the Briilouin function describes how the electric polarization density of an ideal dielectric depends on the external electric field. Hence, it should not be a surprise if the gravitational polarisation density of the quantum vacuum can be approximated by use of a Brillouin function.

The value of $J$ should be related to the number of degrees of freedom of the appropriate virtual constituents of the quantum vacuum. For instance, as we know from quantum field theory, a *massless* spin-1 particle (like photon or gluon) has two degrees of freedom that can be related with $J = 1/2$, effectively the same as if they were fermions. Of course, virtual dipoles with three degrees of freedom cannot be excluded. It is easy to show that

$$\frac{2}{3} < \frac{B_1(x)}{B_{1/2}(x)} < 1 \tag{13}$$

Since the mass of a galactic dark matter halo is not constrained better than a factor of two (Boylan-Kolchin et al. 2013), relation (6) indicates that the present day precision of measurements is not sufficient to distinguish between Brillouin function with $J = 1/2$ and $J = 1$. However it is important to keep in mind that more precise observational findings might provide information about the number of degrees of freedom of virtual gravitational dipoles.

## 3. Discussions

Outside of a spherical body with baryonic mass $M_b$, constants $P_{g\max}$ and $R_c$ can be respectively written (Hajdukovic 2011, 2012) as

$$P_{g\max} = \frac{A}{\lambda_\pi^3}\frac{\hbar}{c}; \quad R_c = \lambda_\pi\sqrt{\frac{M_b}{Bm_\pi}} \tag{14}$$

where $m_\pi$ denotes mass of a pion (typical mass for the physical vacuum in quantum chromodynamics) and $\lambda_\pi = h/m_\pi c$ is the corresponding Compton wavelength; while $A<1$ and $B<1$ are two positive, dimensionless constants of the order of unity (theoretical arguments lead (Hajdukovic 2011, 2012) to $B \approx 2A$ and $A$ is close to the value $1/2\pi$). Let us note that the choice $A = 1/2\pi$ comes from consideration of both dark matter and dark energy (Hajdukovic, 2013) and is slightly different from a previous choice $A \approx 1/2\sqrt{3}$ when only dark matter was considered.

Just as an encouragement for further studies let us mention that according to Boylan-Kolchin et al. 2013, for the Milky Way $M_b \approx 1.3 \times 10^{41} kg$ and the median mass $M_{tot}(260 kpc) = 1.6 \times 10^{12} M_{Sun}$, while our equation (2) with theoretical values for $P_{g\max}$ and $R_c$, gives $M_{dm}(260 kpc) \approx 1.4 \times 10^{12} M_{Sun}$.

## *Appendix: A constant dark matter halo surface density in galaxies*

Recently, using the *Burkert profile* (1), Donato et al. (2009) have concluded that relation $\rho_0 r_0 \approx cons \tan t$ is valid over a range of 14 mag in luminosity and for all Hubble types:

$$\rho_0 r_0 = 140^{+80}_{-30} \frac{M_{Sun}}{pc^2} \tag{15}$$

The work of Donato et al. confirms and extends the previous findings (Kormendy and Freeman 2004, Spano et al. 2008). In the pioneering work, Kormendy and Freeman (2004) have shown that if dark matter distribution is modeled by "*pseudo-isothermal sphere*" the quantity $\rho_0 r_0$ is just below $100 M_{Sun}/pc^2$ (see Fig. 5 in their paper). It is important to note that while we have used the same notation for the free parameters $\rho_0, r_0$ they have *different values* for the Burkert profile and pseudo-isothermal sphere. Spano et al. (2008) have modeled dark matter halo with a cored isothermal sphere and have found that for 36 nearby spiral galaxies $\rho_0 r_0 = 150^{+100}_{-70} M_{Sun}/pc^2$.

Let us rewrite equation (3) in the form

$$\rho_{dm}(r) \cdot r = 2P_{g\max} \left[ \tanh\left(\frac{R_c}{r}\right) - \frac{R_c}{r} \frac{1}{2\cosh^2\left(\frac{R_c}{r}\right)} \right] \tag{16}$$

Evidently, for $r = R_c$, the quantity $\rho_{dm}(R_c) \cdot R_c$ has the same value for all galaxies. More general, $\rho_{dm}(r) \cdot r$ has the same value for different galaxies if $r$ is chosen so that the ratio $R_c/r$ also has the same value for all considered galaxies. Additionally, let us note that expression in brackets in equation (16) changes slowly with $R_c/r$; consequently $\rho(r) \cdot r \approx cons \tan t$ if $r \approx R_c$ or $R_c/r \approx cons \tan t$. This opens possibility to interpret relation $\rho_0 r_0 \approx cons \tan t$ as a signature of the quantum vacuum.

Comparison of results (15) and (16) demands certain caution because there is a significant difference between the right-hand sides of these equations. In (16) we have o product of a distance $r$ and dark matter density $\rho_{dm}(r)$ at that distance. However, in (15) the central density $\rho_0$ is a constant which is different from the dark matter density $\rho_B(r_0) = \rho_0/4$ determined by equation (1).